# Polarization – dependent tunneling of light in gradient optics.

*A. Shvartsburg,[1] V. Kuzmiak [2] and G. Petite [3]*

[1] Science and Technology Center for Unique Instrumentation, Russian Acad. of Sciences, p/o 117342, Butlerov Str. 15, Moscow, Russia.
[2] Institute of Radio Engineering and Electronics, Czech Acad. of Sciences, Chaberska 57, 182 51 Praha 8,  Czech Republic.
[3] Laboratoire des Solides Irradies, CEA – DSM, CNRS, Ecole Polytechnique, 91128, Palaiseau,  France.

**Abstract :** Reflection–refraction properties of photonic barriers, formed by dielectric gradient nanofilms, for inclined incidence of both S- and P-polarized electromagnetic (EM) waves are examined by means of exactly solvable models. We present generalized Fresnel formulae, describing the influence of the non-local dispersion on reflectance and transmittance of single- and double-layer gradient photonic barriers for S- and P- waves and arbitrary angles of incidence. The non-local dispersion of such layers, arising due to a concave spatial profile of dielectric susceptibility across the plane film, is shown to result in a peculiar heterogeneity – induced optical anisotropy, providing the propagation of S(P) waves in tunneling (travelling) regimes. The  results obtained indicate the possibility of narrow–band non–attenuated tunneling (complete transmittance) of oblique S- waves through such heterogeneous barriers, and the existence of spectral areas characterized by strong reflection of P-waves and deep contrast between transmitted S– and P-waves. The scalability of obtained exact analytical solutions of Maxwell equations into the different spectral ranges is discussed and the application potential of these phenomena for miniaturized polarizers and filters  is demonstrated.

# 1.    INTRODUCTION.

Tunneling is a basic wave phenomenon with many interesting applications. The concept of tunneling was pioneered by Gamow for particles in quantum mechanics as long ago as in 1928 [1]. Later on this quantum aproach, involving the time-independent Schrodinger equation, and the wave approach, involving the Helmholtz monochromatic wave equation, were considered to be formally equivalent. The advent of lasers has attracted attention on the effects of tunneling of EM waves through classical barriers in a series of optoelectronic problems, such as, e.g., the evanescent modes in dielectric waveguides [2] and liquid crystals [3], Goos–Hanchen effect for curvilinear interfaces [4-6]. Experiments in the microwave range with "undersized" waveguide [7], and Goos-Hanchen spatial shift of a narrow beam, passing through paraffine double prisms [8], have demonstrated the possibilities of observation of tunneling effects, which proved not to be easy to reproduce at quantum scales. Side by side with the numerous applications of the tunneling concept in optoelectronics, solid state and microwave physics [9], this concept contains an intriguing theoretical challenge, connected with the idea of superluminal light tunneling through the photonic barrier (Hartman paradox, [10], [11]).

The theoretical background of the aforesaid researches is based on the electromagnetics of heterogeneous media [12]. Herein some important features of evanescent waves are examined usually in the framework of a simple model of rectangular opaque barrier [13] or layered structure, formed by step-like variations of refractive indices [14]. Another field for wave tunneling phenomena is connected with the gradient optics of thin films: the non – local dispersion, arising due to continuous spatial variations of dielectric susceptibility was shown recently to provide a rich variety of tunneling phenomena even in a material with negligible natural local dispersion [15]. These phenomena can be examined by means of exact analytical solutions of Maxwell equations for media with spatially distributed dielectric susceptibility

$$\varepsilon(z) = n_0^2 U^2(z).$$  (1)

Here $n_0$ is the value of the host dielectric refractive index, the dimensionless function $U(z)$ describes the coordinate dependence of the dielectric susceptibility. This heterogeneity can result in the formation of a cut-off frequency $\Omega$, dependent upon the gradient and curvature of the $\varepsilon(z)$ profile [15]; thus, for heterogeneous nanofilms with a thickness of about 100 nm and depth of modulation of refractive index about 25% the cut-off frequencies $\Omega$, separating the evanescent and travelling modes, belong to the near IR or visible spectral ranges. Controlled variety of reflection – refraction properties of the layer, produced by different profiles $U(z)$, opens the possibilities of creation of media with optical properties unattainable in natural materials.

 The decisive role of the heterogeneity-induced dispersion was examined in [15] for the case of normal incidence, when the wave polarization effects are vanishing; however, the model $U(z)$ used in [15] does not support analytical solutions of the wave equation for an inclined incidence of polarized waves. Unlike Ref. 15, this paper is devoted to the peculiar effects in reflectivity and transmittivity of both S- and P- polarized waves, incidenting under an arbitrary angle on an heterogeneous dielectric layer, characterized by some profile $U(z)$.

The paper is organized as follows: in Section II we present a model of non-monotonic concave profile $U(z)$, suitable for both S- and P-waves in the gradient medium and demonstrate the effects of heterogeneity – induced optical anisotropy. Exact analytical solutions of the wave equations obtained for such $U(z)$, revealing the polarization-dependent propagation and tunneling regimes for both one and two gradient layers, are examined in Section 3. In section 4 we use these solutions for analysis of narrow – band reflectionless tunneling of S-waves, strong reflection of P-waves, deep contrast between transmitted S- and P- waves and some phase effects, associated with the aforesaid phenomena. The conclusions are summarized in Section 5.

## 2. PROPAGATION AND TUNNELING OF S- AND P–POLARIZED WAVES IN GRADIENT MEDIA (EXACTLY SOLVABLE MODEL).

Unlike in the case of the normal incidence case, here the waves have a different polarization structure and are described by different equations. Denoting the normal to the layer as $z$-axis and choosing the projection of the wave vector on the layer's interface as $y$-axis, one can describe the polarization structure of an S-wave by means of its electric component $E_x$ and magnetic components $H_y$ and $H_z$. The layer is assumed to be lossless and non-magnetic, and one can write the Maxwell equations linking these components:

$$\frac{\partial E_x}{\partial z} = -\frac{1}{c}\frac{\partial H_y}{\partial t} \ ; \qquad \frac{\partial E_x}{\partial y} = \frac{1}{c}\frac{\partial H_z}{\partial t} \ ; \qquad \frac{\partial H_z}{\partial y} - \frac{\partial H_y}{\partial z} = \frac{\varepsilon(z)}{c}\frac{\partial E_x}{\partial t} \ ; \qquad (2)$$

$$div\left(\varepsilon \overline{E}\right) = 0 \ ; \ div\left(\overline{H}\right) = 0 \ . \qquad (3)$$

Components of P-wave ($H_x$, $E_y$ and $E_z$) are also linked by Eq. (3); but the Eqs. (2) have to be replaced by:

$$\frac{\partial H_x}{\partial z} = \frac{\varepsilon(z)}{c}\frac{\partial E_y}{\partial t}; \qquad \frac{\partial H_x}{\partial y} = -\frac{\varepsilon(z)}{c}\frac{\partial E_z}{\partial t} \ ; \quad \frac{\partial E_z}{\partial y} - \frac{\partial E_y}{\partial z} = -\frac{1}{c}\frac{\partial H_x}{\partial t} \ . \qquad (4)$$

The effects of heterogeneity-induced dispersion for inclined incidence can be found from the set (2) – (4) by means of exactly solvable models of $\varepsilon(z)$ given by Eq. (1), suitable for both S- and P- polarizations. The well known Rayleigh profile $U(z) = (1 + z/L)^{-1}$, dating back to 1880 , and exponential profile $U(z) = \exp(z/L)$ provides exact solutions of the set (2) –(4) for monotonic variations of $U(z)$ [12]; however, these models are not suitable for description of smoothly varying concave photonic barriers, that we plan to discuss. Thus, the reflection–refraction problem for the set of equations (2)-(4) is considered below from the very beginning. It is worthwhile to express the field components in Maxwell equations by means of the following auxiliary, polarization-dependent functions $\Psi_s$ and $\Psi_p$ :

S – polarization: $E_x = -\dfrac{1}{c}\dfrac{d\Psi_s}{dt}$ ; $\qquad H_y = \dfrac{d\Psi_s}{dz}$ ; $\quad H_z = -\dfrac{d\Psi_s}{dy}$ ; $\qquad\qquad$ (5)

P – polarization: $H_x = \dfrac{1}{c}\dfrac{d\Psi_p}{dt}$ ; $\qquad E_y = \dfrac{1}{\varepsilon(z)}\dfrac{d\Psi_p}{dz}$ ; $\qquad E_z = -\dfrac{1}{\varepsilon(z)}\dfrac{d\Psi_p}{dy}$ . $\qquad$ (6)

Using such representations one can reduce the system (2) – (4) to two equations, governing S- and P-waves respectively. Omitting for simplicity the factor $\exp[i(k_y y - \omega t)]$, these equations can be written as:

$$\frac{\partial^2 \Psi_s}{\partial z^2} + \left(\frac{\omega^2 n_0{}^2 U^2}{c^2} - k_y{}^2\right)\Psi_s = 0 \ , \qquad k_y = \frac{\omega}{c}\sin\theta \ , \qquad\qquad (7)$$

$$\frac{\partial^2 \Psi_p}{\partial z^2} + \left(\frac{\omega^2 n_0^2 U^2}{c^2} - k_y^2\right)\Psi_p = \frac{2}{U}\frac{dU}{dz}\frac{\partial \Psi_p}{\partial z} \ . \qquad\qquad (8)$$

By introducing the new variable $\eta$ and new functions $f_s$ and $f_p$

$$\eta = \int_0^z U(z_1)dz_1 \; ; \; f_s = \frac{\Psi_s}{\sqrt{U(z)}} \; ; \qquad f_p = \Psi_p\sqrt{U(z)} \; ; \qquad (9)$$

one can present eqs. (7) – (8) for S- and P – waves in the forms:

$$\frac{d^2 f_s}{d\eta^2} + f_s\left(\Lambda - \frac{U_{\eta\eta}}{2U} + \frac{U_\eta^2}{4U^2}\right) = 0 \qquad\qquad (10)$$

$$\frac{d^2 f_p}{d\eta^2} + f_p\left(\Lambda + \frac{U_{\eta\eta}}{2U} - \frac{3U_\eta^2}{4U^2}\right) = 0 \; , \qquad\qquad (11)$$

where $\Lambda = \dfrac{\omega^2 n_0^2}{c^2} - \dfrac{k_y^2}{U^2}$ ; $\qquad U_\eta = \dfrac{dU}{d\eta}$ ; $\quad U_{\eta\eta} = \dfrac{d^2 U}{d\eta^2}$ . Eqs. (10) – (11) are valid for

arbitrary profiles of photonic barriers $U(z)$ and all angles of incidence $\theta$, including in particular, the two well – known exactly solvable models mentioned above (Rayleigh and exponential profiles). The fact that some polarization effects will depend on the barrier profile is clear even without solution of these equations. In the case of the widely used Rayleigh profile $U_R$, which can be rewritten in $\eta$-space by means of (9) as $U_R(\eta) = \exp(-\eta/L)$, therefore both equations (10) and (11) coincide:

$$\frac{d^2 f_{s,p}}{d\eta^2} + f_{s,p}\left(\frac{\omega n_0}{c}\right)^2\left(1 - \frac{\omega_{cr}^2}{\omega^2} - \frac{\sin^2\theta}{n_0^2}\exp\left(\frac{2\eta}{L}\right)\right) = \mathbf{0.} \qquad\qquad (12)$$

Eq. (12) shows that, due to heterogeneity-induced dispersion, characterized by a cut-off frequency $\omega_{cr} = c/2n_0 L$, the tunneling regimes arise in the Rayleigh barrier for both S- and P-waves simultaneously. In the case of the exponential profile, rewritten in $\eta-$space as $U = 1 - \eta/L$, no heterogeneity -induced cut-off frequency can be defined from eqs. (10)-(11). In contrast to these traditionally used asymmetrical profiles, we consider here the exactly solvable symmetrical concave barrier $U(z)$, formed inside the dielectric film (thickness $d$) by variation of $U$ from the value $U = 1$ in the centre of barrier $(z = 0)$ up to the values $U_{m}$ at the interfaces $z = \pm d/2$ (Fig. 1):

$$U = \frac{1}{\cos\left(\dfrac{z}{L}\right)}; \qquad U_m = \frac{1}{m}; \qquad m = \cos\left(\frac{d}{2L}\right). \tag{13}$$

Note that this model is characterized by one free parameter (the length-scale $L$), however it can be used to study many features associated with this type of potential as we will show in Section 4. In fact, it represents a particular case of a more general potential with two adjustable parameters which, however, leads to much more cumbersome calculations and thus is not developed here. Substitution of (13) into (9) brings the variable $\eta$ and profile U($\eta$) :

$$\eta = L \ln\left[\frac{1 + tg\left(z/2L\right)}{1 - tg\left(z/2L\right)}\right]; \qquad U = \frac{1}{\cos\left(\dfrac{z}{L}\right)} = ch\xi; \;\; \xi = \frac{\eta}{L}. \tag{14}$$

Using eq. (14) one can rewrite eqs. (10)-(11) for S- and P-waves inside the barrier given by eq. (13) into the similar form:

$$\frac{d^2 f_{s,p}}{d\xi^2} + f_{s,p}\left(q^2 - \frac{T_{s,p}}{ch^2\xi}\right) = 0 \tag{15}$$

$$q^2 = \left(\frac{\omega n_0 L}{c}\right)^2 \left(1 - \frac{\Omega^2}{\omega^2}\right) \; ; \tag{16}$$

$$T_s = (k_y L)^2 + \frac{1}{4} \; ; \qquad T_p = (k_y L)^2 - \frac{3}{4} \; , \tag{17}$$

where $\Omega$ is a characteristic frequency, determined by the parameters of concave barrier (13):

$$\Omega = \frac{c}{2n_0 L} = \frac{c}{n_0 d} Arc\cos(m) \; . \tag{18}$$

The sign of parameter $q^2$ given by eq.(16) changes in $\omega = \Omega$. Let us consider here the low frequency spectral range $\omega < \Omega$, $q^2 < 0$. In this range the expression in brackets in eq.(15) for S-wave is always negative, and thus low frequency S-wave, incidenting on the barrier given by eq.(13) under an arbitrary angle $\theta$, is traversing this barrier as an evanescent wave.

On the contrary, the same expression in Eq. (15) for a P-wave, for realistic values of the modulation of the refractive index (13) in the layer ($U_m = m^{-1} \leq 1.3 - 1.4$) remains positive and provides the travelling mode regime for P-wave.

Thus, unlike the homogeneous rectangular barrier, where the tunneling of EM waves is determined by a condition common to both S- and P-polarizations, the tunneling through the concave barrier given by eq.(13) proves to be polarization – dependent. This heterogeneity – induced anisotropy can result in a significant difference in reflectivity and transmittivity of such barrier for S- and P-waves, their frequencies and angles of incidence on the film being equal. This difference is illustrated below by means of exactly solvable model (13).

## 3. HETEROGENEITY–INDUCED OPTICAL ANISOTROPY OF TRANSPARENT LAYERS.

To simplify the analysis of the transmission of a concave photonic barrier for an EM wave with inclined incidence, let us first consider one layer without substrate embedded in free space - see Fig. 1. We introduce the dimensionless quantities

$$u = \Omega/\omega > 1 \; ; \qquad N = \sqrt{1 - u^{-2}} < 1 \,. \tag{19}$$

By transforming eq.(15) by means of new variable $\nu$ and new function $W$

$$v = \frac{1 - th\xi}{2} \; , \; f_{s,p} = (ch\xi)^{\frac{N}{2}} W_{s,p} \tag{20}$$

one obtains for this function the hypergeometric equation

$$v(1-v)\frac{d^2W}{dv^2} + \left[\gamma - v(1+\alpha+\beta)\right]\frac{dW}{dv} - \alpha\beta W = 0 \; ; \; \gamma = 1 - \frac{N}{2} \,. \tag{21}$$

Although eq.(21) is valid for both polarizations, the values of parameters $\alpha$ and $\beta$ have to be specified for each wave; definition $\gamma$ (21) is valid for both waves. Let us start the analysis from S- wave; in this case

$$\alpha_s, \beta_s = \frac{1}{2}\left[1 - N \pm \frac{i\sin\theta\sqrt{1-N^2}}{n_0}\right]. \tag{22}$$

Since $\alpha_s + \beta_s + 1 = 2\gamma$, two linearly-independent solutions of eq.(21) are given by the hypergeometric functions $F(\alpha_s, \beta_s, \gamma, \nu)$ and $F(\alpha_s, \beta_s, \gamma, 1-\nu)$, denoted below for compactness as $F(\nu)$ and $F(1\text{-}\nu)$; moreover, due to condition $\text{Re}(\alpha_s + \beta_s - \gamma) < 0$ the series, presenting these functions, are converging absolutely [16]. The general solution of eq.(21) is

$$W = F(\nu) + Q_s F(1\text{-}\nu) .$$  (23)

Here $F(\nu)$ and $F(1\text{-}\nu)$ can be considered as forward and backward waves (meaning more exactly in a tunneling case, evanescent and antievanescent waves), meanwhile the factor $Q_s$ is associated with the contribution of backward wave to the complete field. By using the variables defined in Eqs. (9) and (19) and expressing the factor $\text{ch}\,\xi$ in terms of $\cos(z/L)$ according to eq.(14) one can present the generating function $\Psi_s$ in a form:

$$\Psi_s = \left[\cos(z/L)\right]^{\frac{1-N}{2}} W .$$  (24)

Substitution of eq.(24) into eq.(5) yields the explicit expressions for components of S-wave inside the medium:

$$E_x = \frac{i\omega A}{c}\,\Psi_s ; \qquad H_z = -ik_e A \Psi_s ;$$

(25)

$$H_y = \frac{-A}{2L\left[\cos(z/L)\right]^{\frac{N+1}{2}}}\left\{(1-N)\sin(z/L)W + \cos^2(z/L)\frac{dW}{d\nu}\right\},$$

where $A$ is the wave's complex amplitude. To find the reflection coefficient for S-wave $R_s$ one has to use the continuity conditions on the film interfaces $z = \pm d/2$. It follows from (14), that

$$th\xi = \sin(z/L) \; ; \qquad v = \frac{1 - \sin(z/L)}{2} \; ; \qquad 1 - v = \frac{1 + \sin(z/L)}{2} \; . \tag{26}$$

Let us consider the wave incidenting on the interface $z = -d/2$. By introducing the variables $v_{1,2}$

$$v\Big|_{z=-\frac{d}{2}} = v_1 = \frac{1+s}{2} \; ; \; (1-v)\Big|_{z=+\frac{d}{2}} = v_2 = \frac{1-s}{2} \; ; \; s = \sqrt{1-m^2} \; ;$$
$$\tag{27}$$

$$F_{1,2} = F(v_{1,2}) \; ; \; F_1' = \frac{dF}{dv}\Big|_{v_1} \; ; \; F_2' = \frac{dF}{dv}\Big|_{v_2}$$

one can derive the expression for $R_s$ from the continuity conditions on this plane

$$R_s = \frac{M_1 + i\zeta_s F_1 + Q_s(M_2 + i\zeta_s F_2)}{i\zeta_s F_1 - M_1 + Q_s(i\zeta_s - M_2)} \tag{28}$$

$$M_{1,2} = s(N-1)F_{1,2} \pm m^2 F_{1,2}' \quad \zeta_s = \frac{m\cos\theta\sqrt{1-N^2}}{n_0} \; . \tag{29}$$

The factor $Q_s$ can be determined from the continuity conditions on the opposite interface $z = d/2$ :

$$Q_s = -\left(\frac{M_2 - i\zeta F_2}{M_1 - i\zeta F_1}\right). \qquad (30)$$

Substituting $Q_s$ given by eq.(30) into eq.(28) one obtains the complex reflection coefficient $R_s$

$$R_s = \frac{A_s}{B_s} \; ; \; A_s = M_1^2 - M_2^2 + \zeta_s^2\left(F_1^2 - F_2^2\right); \; B_s = B_1 + iB_2 \; :$$

$$(31)$$

$$B_1 = M_2^2 - M_1^2 + \zeta_s^2\left(F_1^2 - F_2^2\right) \; ; \; B_2 = 2\zeta_s\left(F_1 M_1 - F_2 M \right).$$

Now we rewrite the complex reflection coefficient $R_s$ into the form $R_s = |R_s|\exp(i\phi_{sr})$

$$|R_s| = \frac{|A_s|}{\sqrt{B_1^2 + B_2^2}} \; ; \; \phi_{sr} = -arctg\left(\frac{B_2}{B_1}\right). \qquad (32)$$

To find the transmission function $T_s$ one has to express the amplitude of refracted wave $A$ from (25) via the amplitude of incidenting wave $E_0$ and reflection coefficient $R_s$

$$A = \frac{-iE_0 c\left(1 + R_s\right)}{\omega\left(F_1 + Q_s F_2\right)} \qquad (33)$$

By using the eq.(33) one can write the field $E$ at the interface $z = d/2$ by means of a complex transmission function $T_s$:

$$E = E_0 T_s \; ; \qquad T_s = |T_s|\exp\left(i\phi_{st}\right) \; ; \qquad |T_s| = \frac{2\zeta_s|F_1 M_2 - F_2 M_1|}{\sqrt{B_1^2 + B_2^2}}; \qquad \phi_{st} = arctg\left(\frac{B_1}{B_2}\right). \qquad (34)$$

The reflection coefficient for P-wave $R_p$ can be calculated in the same way. The generating function $\Psi_p$ (6) is expressed via the relevant solution of the eq.(21), similar to eq.(23):

$$\Psi_p = \left[\cos(z/L)\right]^{-\frac{N+1}{2}} \left[F\left(\alpha_p, \beta_p, \gamma, \nu\right) + Q_p F\left(\alpha_p, \beta_p, \gamma, 1-\nu\right)\right];$$

$$Q_p = -\left(\frac{M_4 - i\zeta_p F_4}{M_3 - i\zeta_p F_3}\right); \tag{35}$$

$$\zeta_p = \frac{n_0 \cos\theta \sqrt{1-N^2}}{m};$$

Forward and backward(propagating) waves are described through the eq.(35) by the hypergeometric functions $F$ with parameters

$$\alpha_p, \beta_p = \frac{1}{2}\left[1 - N \pm \sqrt{4 - \frac{\left(1-N^2\right)\sin^2\theta}{n_0^2}}\right]; \quad \gamma = 1 - \frac{N}{2}; \tag{36}$$

Substitution of generating function given by eq.(35) into eq.(6) brings the components of the P-wave. Using the continuity conditions, one can present the reflection coefficient $R_p$ in a form, analogous to $R_s$ (32) :

$$R_p = \frac{M_3 + i\varepsilon_p F_3 + Q_p\left(M_4 + i\varepsilon_p F_4\right)}{i\varepsilon_p F_3 - M_3 + Q_p\left(i\varepsilon_p F_4 - M_4\right)};$$

$$R_p = |R_p|\exp\left(i\phi_{pr}\right) \quad |R_p| = \frac{|A_p|}{\sqrt{B_3^2 + B_4^2}}; \quad \phi_{pr} = -arctg\left(\frac{B_4}{B_3}\right);$$

$$A_p = M_3^2 - M_4^2 + \zeta_p^2\left(F_3^2 - F_4^2\right) \; ; \tag{37}$$

$$B_3 = M_4^2 - M_3^2 + \zeta_p^2\left(F_3^2 - F_4^2\right) \; ; \qquad B_4 = 2\zeta_p\left(F_3 M_3 - F_4 M_4\right) \; ;$$

$$M_{3,4} = s(N+1)F_{3,4} \pm m^2 F_{3,4}^{'} ; \qquad F_{3,4}^{'} = \frac{dF\left(\alpha_p, \beta_p, \gamma, v\right)}{dv}\Big|_{v=v_{1,2}} \; .$$

The transmission function for P-wave $T_p$ can be found by analogy with eqs.(33) – (34) in the form

$$T_p = \left|T_p\right|\exp\left(i\phi_{pt}\right) \; ; \quad \left|T_p\right| = \frac{2\zeta_p\left|F_3 M_4 - F_4 M_3\right|}{\sqrt{B_3^{23} + B_4^2}} \; ; \quad \phi_{pt} = arctg\left(\frac{B_3}{B_4}\right) \; ; \tag{38}$$

Thus, we found the expressions for reflection coefficients $R_{s,p}$ and transmission functions $T_{s,p}$, valid for an arbitrary angle of incidence $\theta$. The quantities $|R_{s,p}|$ and $|T_{s,p}|$ are linked by the energy conservation law:

$$|T_{s,p}|^2 = 1 - |R_{s,p}|^2 \; . \tag{39}$$

It is necessary to emphasize that in the normal incidence case ($\theta = 0$) the difference between S- and P- polarized waves is vanishing, and, thus, the moduli of reflection and transmission coefficients have to be equal; this equality, which is not obvious from the formulae for $|R_s|$ (32) and $|R_p|$ (37), is demonstrated analytically in the Appendix.

The effects of heterogeneity-induced dispersion on the amplitude-phase structure of transmitted radiation manifest themselves, first of all, by the angular dependence of the transmittance of gradient layers for S– and P–polarized waves. These transmittances, $|T_s|^2$ and $|T_p|^2$, calculated by eqs.(34) and (38) respectively, are depicted on Fig. 2. In the case of inclined incidence $|T_p|^2$, exceeding $|T_s|^2$, can even reach the value $|T_p|^2=1$, illustrating the Brewster effect for a gradient layer; however, the angle $\theta$, related to this reflectionless regime, differs from the Brewster angle $\theta_B$ for an homogeneous layer ($tg\,\theta_B = n_0$).

In Fig. 2 we present the angular dependence of transmittances of an heterogeneous layer for S- and P- waves for some given value of dimensionless parameter $\gamma$ given by eq.(21), defined for any normalized frequency $u = \Omega/\omega$. Considering the transmittance regime, related to some values $\gamma$, m, $\theta$ and $n_0$ and using the characteristic frequency $\Omega$ given by eq.(18), one can choose the thickness of the layer $d$, providing the formation of such regime for any given polarization and wavelength $\lambda$ by means of expression, following from eqs.(21) and (18):

$$\frac{d}{\lambda} = \frac{\sqrt{1-4(1-\gamma)^2}\, Arc\cos(m)}{2\pi n_0}.$$  (40)

Thus, for the set of parameters related to Fig. 2 and $\lambda = 1.55\,\mu m$ eq. (40) defines the thickness of the layer $d = 45$ nm; herein Fig. 2 remains valid for any values of $d$ and $\lambda$, linked by relation (40) with $m = 0.75$, $n_0 = 3.4$ and $\gamma = 0.75$.

To compare these results with the transmittance of an homogeneous layer ($m = 1$) with the same values of $d$ and $n_0$, one can examine the simple case of normal incidence. The reflection coefficient $R$ for this geometry is known [19] to be given by

$$R = \frac{(1-n_0^2)tg\,\delta}{(1+n_0^2)tg\,\delta + 2in_0}; \qquad \delta = \frac{2\pi n_0 d}{\lambda}.$$  (41)

Calculation of $R$ for the abovementioned values of $d$, $n_0$ and $\lambda$ and substitution of this $R$ into the conservation law (39) yields the transmittance of the homogeneous layer for normal incidence $|T_0|^2 = 0.33$, meanwhile the transmission coefficient for the heterogeneous layer shown in Fig. 2, for normal incidence is $|T|^2 = 0.46$. Thus, the transmittance of a gradient layer $|T|^2$ can exceed the transmittance of an homogeneous film $|T_0|^2$ with the same values of $d$ and $n_0$ by 40%.

The spectral properties of transmittance for S- and P- waves are presented in Figs. 3-4 and 6 in the ($|T|^2 - \gamma$) plane, providing a uniform scale on horizontal axis of graphs. Herein the comparison of Figs. 3a – 3b, drawn for one gradient layer, shows, that the increase of depth of modulation of refractive index $m$ can result in drastic changes of transmittance – reflection spectra of gradient film for S-waves. A narrow asymmetrical peak of reflectionless tunneling for S-waves arises near by the point $u = 1$, $\gamma = 1$. This peak is contiguous with a narrow area of high dispersion of transmittance coefficient with almost vertical tangent to the graph $|T_s(\gamma)|^2$. The transmittance $|T_p|^2$ in this range remains almost constant, approximately 87 %. When the angle of incidence is decreased the existence of the peak is unaffected while transmittance $|T_p|^2$ tends to 100% (Fig. 3c).

On the contrary, in the low frequency spectral range, close to the point $\gamma = 0.5$ (Fig.3c), the reflection of S- waves is almost unvariable, close to 100%, meanwhile the transmittance of P-waves tends to zero; herein the frequency dispersion of $|T_p|^2$ in this range is strong. This high contrast between the transmittance of gradient layers for S- and P-waves may be interesting for polarizing systems, operating under the large angles of incidence. These layers may be rather thin: thus, e.g., a polarizing screen, providing for transmitted waves the ratio $|T_p|^2/|T_s|^2 < 0.05$, is characterized, under the conditions of Fig. 3c ($u = 3.2$), by the ratio $d/\lambda = 0.025$ (eq.40). Such miniaturized scales $d$ are a remarkable feature of the anisotropic gradient nanolayers in question.

## 4. NARROW-BAND REFLECTIONLESS TUNNELING OF S –WAVES.

While discussing the transmittance peaks for S-waves $|T_s|^2 = 1$ in the spectral range $u \leq 1$ (Fig. 3b – 3c), one has to emphasize that these peaks arise in the regime of reflectionless tunneling of wave ($R_s = 0$) through the gradient layer. Cancellation of the reflected wave results from the interference of the wave reflected on the interface $z = -d/2$, with the transmitted part of backward antievanescent wave. This cancellation arises only for the concave photonic barrier: in the case of a square barrier with constant refractive index such cancellation proves to be impossible [17].

Proceeding in a similar fashion, we can find the transmittance of a pair of gradient layers for S- and P – polarized waves, characterized by coefficients $|T_{2s}|^2$ and $|T_{2p}|^2$. To calculate these coefficients let us examine the set of two parallel adjacent layers, shown in Fig. 1. Herein the continuity conditions on the interfaces $z = -d/2$ and $z = 3d/2$ remain unchanged. Considering first the S-wave one can see, that formulae for $R_s$ (28) and $Q_s$ (30) relating to these conditions for S-wave, are also valid for the set of two layers. Recalling these conditions for the intermediate boundary $z = d/2$, we can find, after some tedious algebra, the value $Q$, related to this boundary

$$Q = -\left[ \frac{F_1 M_2 + M_1 F_2 + 2Q_s M_2 F_2}{Q_s (F_1 M_2 + M_2 F_1) + 2M_1 F_1} \right] . \qquad (42)$$

Here the values $M_{1,2}$ and $Q_s$ are defined in eqs.(29) – (30). Substitution of eq.(42) into eq(28) yields the complex reflection coefficient $R_{2s}$ for the set of two layers:

$$R_{2s} = \frac{G_s}{K_s + iJ_s}$$

$$G_s = (F_1 M_1 - F_2 M_2)[M_1^2 - M_2^2 + \zeta_s^2 (F_1^2 - F_2^2)] ;$$

$$K_s = ( F_1M_1 - F_2M_2 )[ M_2^2 - M_1^2 + \zeta_s^2 (F_1^2 - F_2^2)] \;;$$

$$J_s = \zeta_s [ (F_1^2 - F_2^2)(M_1^2 - M_2^2) + (F_1M_1 - F_2M_2)^2] \;.$$

The reflection coefficient of the same pair for P-wave $R_{2p}$ can be determined by analogy with eq.(43) by means of replacement of indices $s \rightarrow p$ in the relevant terms in eqs.(42) – (43), e.g., $G_s \rightarrow G_p$, and the following transpositions:

$$R_{2p} = \frac{G_p}{K_p + iJ_p} \;;$$

(44)

$$F_{1,2} \rightarrow F_{3,4} \;;\; M_{1,2} \rightarrow M_{3,4} \;;\; \zeta_s \rightarrow \zeta_p \;;$$

In the case of normal incidence one again obtains $|R_{2s}|^2 = |R_{2p}|^2$ - see Appendix.

The transmission coefficients $|T_{2s}|^2$ and $|T_{2p}|^2$ for the pair of layers, found by means of the substitution of eq.(43) and eq.(44) into eq.(39), are presented on Fig. 4. The comparison of Figs. 4a and 4b shows the strengthening of optical anisotropy of gradient films due to increase of their heterogeneity: The decrease of $m$ from $m = 0.95$ (Fig.4a) to $m= 0.86$ (Fig. 4b) results again in the formation of reflectionless tunneling (non-attenuated transmittance) regimes for S-waves as well as the huge dispersion of $|T_{2s}|^2$ and slowly varying high transmittance of P-waves near by the point $u = 1$. However, in this geometry the range of reflectionless tunneling contains two closely located peaks with the very narrow spectral width (Fig. 5). Thus, such pair of films can be considered as a model of miniaturized frequency filter for S-waves. The effects of non-attenuated tunneling analogous to superlensing phenomenon in which evanescent waves contribute to the perfect image of the objects by means of negative refractive index medium[18], represent an alternative and new concept of energy transfer that employs evanescent waves and may be useful in design of subwavelength devices.

In Fig. 6 we present the phase shifts of transmitted waves $\phi$ . The phase times $t_{\text{ph}} = \dfrac{\partial \phi}{\partial \omega}$ are positive for S-wave in all the spectral range $u < 1$, meanwhile for P-wave the values $t_{\text{ph}}$ are positive in a broad spectral interval $0.55 < \gamma < 1$ and negative in a narrow spectral interval $0.52 < \gamma < 0.55$. In the former interval the phase shift of P- waves, passing through the pair of films, can reach the values, close to $\pm \pi/2$.

Comparison of phases of reflected (32) and transmitted (34) S- waves leads to the relation $tg\,\phi_{sr}\,tg\,\phi_{st} + 1 = 0$. A similar relation can be found from eqs.(37) – (38) for P-waves also. Thus, these phases are linked:

$$\phi_{sr} - \phi_{st} = \phi_{pr} - \phi_{pt} = \pm \frac{\pi}{2} \tag{45}$$

Formulae (43)-(44) show that the correlation (45), derived for one layer, remains valid for a pair of layers. This property can be used for determination of the phase of tunneling wave, if its detection is impeded due to strong attenuation in an opaque barrier [19].

Until now our analysis was restricted by the model of heterogeneous film without substrate. To examine the applicability of obtained results to the real case of a layer supported by substrate, let us consider the layer deposited on an homogeneous lossless layer of thickness $D$ and with refractive index $n_D$ – see Fig.1 . By presenting the generating function for, e.g., S-wave, inside the homogeneous layer in a form

$$\psi_D = \left[\exp(ik_\perp z) + Q_D \exp(-ik_\perp z)\right]\exp\left[i\left(k_y y - \omega t\right)\right] ;$$

$$\tag{46}$$

$$k_\perp = \frac{\omega}{c}r ; \ r = \sqrt{n_D^2 - \sin^2 \theta} \ ,$$

and by using the continuity conditions on the interface between this layer and air ($z = d + D$), one can find the parameter $Q_D$ that enters eq.(46) in the form:

$$Q_D = -\left(\frac{\cos\theta - r}{\cos\theta + r}\right)\exp\left(2ik_\perp D\right).$$

(47)

The continuity conditions on the interface between the layer and heterogeneous film can be written by means of eqs.(25)–(27)

$$\frac{M_2 + Q_s M_1}{F_2 + Q_s F_1} = \frac{2i\omega Lrm}{c}\left(\frac{1 - Q_D}{1 + Q_D}\right).$$

(48)

The expression in brackets in (48) can be rewritten by means of eq.(47) in the form

$$\frac{1 - Q_D}{1 + Q_D} = \frac{\cos\theta - i\,r\,tg\,\zeta}{r - i\cos\theta}\;;\quad \zeta = \frac{\omega r D}{c};$$

(49)

If the thickness $D$ is choosen so, that

$$\frac{\omega D\sqrt{n_D^2 - \sin^2\theta}}{c} = l\pi\;;\qquad l = 1;2;3...$$

(50)

where the right side of eq.(48) is reduced to $i\zeta_s$ (29), and the value $Q_s$, defined from (48), as well as the reflection coefficient $R_s$ coincide with the values (30) and (32), calculated in the absense of substrate. The same condition (50) can be also found for P-wave. Thus, the condition (50) being fulfilled, this layer makes no influence on the reflection-refraction properties of gradient films discussed.

When the gradient layer is deposited on a homogeneous transparent substrate with refractive index $n$ and thickness $D$, the expressions for reflection coefficients $R_s$ and $R_p$ obtained above can be generalized by means of the relevant continuity conditions on the interface z = d/2. In particular, in a case, when $D >> ct$, where t is the duration of incident wave pulse (thick substrate) – so that no interference effects can occur between the incident and the far-interface reflected pulse – or when the reflection is eliminated by use of an anti-reflection coating or by using a wedged substrate, these generalized expressions for a single layer read as:

$$R_s = \frac{A_{s1}}{B_{s1}} \; ; \quad A_{s1} = M_1^2 - M_2^2 + \zeta_s \zeta_1 \left(F_1^2 - F_2^2\right) + i\left(\zeta_s - \zeta_1\right)\left(F_1 M_1 - F_2 M_2\right) \; ;$$

$$B_{s1} = M_2^2 - M_1^2 + \zeta_s \zeta_1 \left(F_1^2 - F_2^2\right) + i\left(\zeta_s + \zeta_1\right)\left(F_1 M_1 - F_2 M_2\right) \; ; \quad \zeta_1 = \zeta_s \frac{r}{\cos\theta} \; ; \tag{51}$$

$$r = \sqrt{n^2 - \sin^2\theta} \; .$$

The value $R_p$ can be obtained from this $R_s$ by applying the transpositions:

$$A_{s1} \rightarrow A_{p1}; B_{s1} \rightarrow B_{p1} \; ; F_{1,2} \rightarrow F_{3,4} \; ; M_{1,2} \rightarrow M_{3,4} \; ; \zeta_s \rightarrow \zeta_p \; ; \zeta_1 \rightarrow \zeta_p \frac{r}{\cos\theta}.$$

In a case when n = 1 (air) and r = $\sin\theta$, these formulae reduce to eqs.(31) and (37), respectively. The condition of reflectionless tunneling $R_{1s} = 0$ results in two equations, nullifying the real and imaginary parts of numerator $A_{s1}$. Combining the equations $Re[A_{s1}] = 0$ and $Im[A_{s1}] = 0$, one obtains the expression $r$:

$$\frac{r}{\cos\theta} = \frac{M_2^2 - M_1^2}{\zeta_s^2 \left(F_1^2 - F_2^2\right)} \; . \tag{52}$$

The right side of eq.(52) is known, therefore $r = \sqrt{n^2 - \sin^2\theta}$ and, finally, one obtains the value of the refractive index of substrate $n$, that provides such non–attenuated tunneling. Then the values $T_s$ and $T_p$ can be calculated using eq. (39). An example of transmittance through a gradient photonic barrier deposited on a thick substrate with refractive index $n$ for S and P wave, calculated according to eq.(52), is depicted in Fig. 7 and denoted by s and p, respectively. We find that the presence of the thick substrate affects the transmittance for S wave (denoted by $S_0$ in Fig. 7) and results in broadening of the peak of the transmittance and an increase of its minimum in comparison with the transmittance without the substrate, however, the main tendencies in spectrum remain unchanged, while the change of the transmittance for P wave is negligible.

## 5. CONCLUSIONS.

In conclusion, we have considered the transmittance of gradient photonic barriers, formed by thin dielectric layers with concave profiles of refractive index, for arbitrarily polarized EM waves, incidenting on these barriers under arbitrary angles.. The non–local dispersion, determined by the shape of photonic barriers, is shown to provide a peculiar optical anisotropy, stipulating the propagation of P (S)–polarized waves in travelling (tunneling) regime. The amplitude-phase structure of reflected and transmitted S- and P–polarized waves is found in the framework of exactly solvable model of gradient barriers, and the generalized Fresnel formulae for reflectance and transmittance of single layer and double-layer concave photonic barriers are presented. The effect of narrow – band reflectionless tunneling (100 % transmittance) of S–wave is demonstrated. These solutions, obtained without any assumptions regarding the smallness or slowness of variations of EM fields or media, can be used for applicability check of some approximations, found by transfer matrix approach [20] or numerical solutions for EM fields in spatially varying media [21]. The examples of using of these results in the gradient optics of nanolayers may become useful for design of miniaturized optoelectronic devices with dimensions much smaller than the wavelength, operating with the oblique incidence of EM waves, such as polarizers, phase shifters, frequency-selective interfaces and large incidence angles filters. The scalability of obtained exact analytical solutions to the different spectral ranges seems to be perspective for creation of materials with the electromagnetic properties, unattainable in the natural media.

## ACKNOWLEDGEMENTS.

One of the authors (A.S.) acknowledges the stimulating discussions with Prof. T. Arecchi and A. Migus. A.S and G.P. acknowledge the support of NATO (grant n° PST.CLG 980331) . Work of V.K. was partially supported by COST P11 Action.

## APPENDIX

To confirm the equality of reflection coefficients $|R_s|^2$ and $|R_p|^2$ for the normal incidence ($\theta = 0$) one can use the general properties of hypergeometric functions F [16]:

$$F(\alpha, \beta, \gamma, v) = 1 + \frac{\alpha\beta}{\gamma}v + ...$$

$$\frac{dF}{dv} = \frac{\alpha\beta}{\gamma}F(\alpha+1, \beta+1, \gamma+1, v) \; ; \tag{A1}$$

$$v(1-v)\frac{dF(\alpha, \beta, \gamma, v)}{dv} = (\gamma - \alpha)F(\alpha-1, \beta, \gamma, v) - (\gamma - \alpha - \beta v)F(\alpha, \beta, \gamma, v) \; .$$

The values of parameters $\alpha$ and $\beta$ for S and P waves for the normal incidence are found from (22) and (36) respectively:

$$\alpha_s = \beta_s = \frac{1-N}{2} \; ; \qquad \alpha_p = \frac{3-N}{2} \; ; \; \beta_p = -\left(\frac{1+N}{2}\right) \; . \tag{A2}$$

The product $v(1-v)$ in (A1), calculated for the planes $z = \pm d/2$, is equal to $m^2/4$. Parameters $M_{1,2}$ (29) and $M_{3,4}$ (37) can be rewritten by means of (A1)–(A2) as

$$M_1 = (1 - N) \, F_3 \; ; M_2 = - (1 - N) \, F_4 \; ;$$

and

$$\tag{A3}$$

$$M_3 = - (1 + N) \, F_1 \; ; M_4 = (1 + N) \, F_2 \; .$$

Substitution of (A3) into the formula for $R_s$ (31) and comparison with $R_p$ (37) yields finally the physically clear result for normal incidence on the gradient layer :

$$R_p = -R_s \; ; \; |T_s|^2 = |T_p|^2 \; . \tag{A4}$$

The same substitution into expression (43) for $R_{2s}$ yields the similar results for normal incidence on a pair of layers:

$$R_{2p} = -R_{2s} \; ; \; |T_{2s}|^2 = |T_{2p}|^2 \; . \tag{A5}$$

**REFERENCES.**

**Figure Captions**

*Figure 1* : Set of gradient layers, supported by an homogeneous substrate; thickness of each layer and substrate are $d$ and $D$ respectively; profile of refractive index inside each layer is given in (13).

*Figure 2* : Transmittance of gradient layer with parameters $m = 0.75$, $n_0 = 3.4$ for S- and P- polarized waves vs. the angle of incidence $\theta$ , $\gamma(\omega) = 0.75$.

*Figure 3* : Spectra of transmittance of single layer ($n_0 = 1.4$) for inclined incidence of S- and P- waves, vs the frequency – dependent parameter $\gamma$ (21); (a): $m = 0.95$, $\theta = 75^0$; (b): $m = 0.75$, $\theta = 75^\circ$ ;(c): $m = 0.75$, $\theta = 65^\circ$ .

*Figure 4 :* Spectra of transmittance of a pair of layers ($n_0 = 1.4$, $\theta = 75^\circ$) for S- and P- waves. (a): $m = 0.95$ ;(b): $m = 0.86$.

*Figure 5 :*  Narrow–band peaks of reflectionless  tunneling of S- waves through the pair of gradient layers ( $n_0 = 1.4$, $m = 0.75$, $\theta = 65^\circ$ ) : The transmission coefficients are plotted vs the normalized frequency $u$.

*Figure 6* : Phases of  S and P waves, passing through the pair of layers under the conditions, given in the Caption to Fig. 4(a).

*Figure 7*:  Transmittance of S and P waves through one layer gradient photonic barrier, deposited on the thick substrate with n = 2.32 (m = 0.86, $\theta = 65^o$, n 0 = 1.4); curve $S_0$ shows the transmittance of the same barrier without substrate.

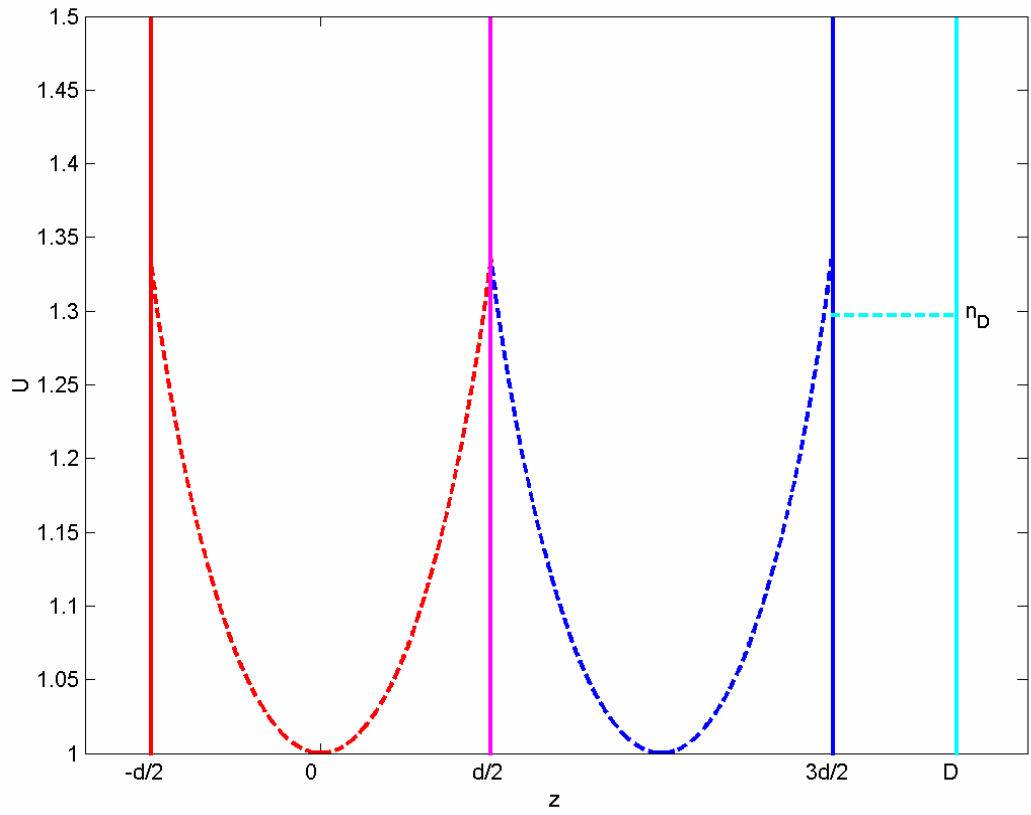

**FIGURE 2 :**

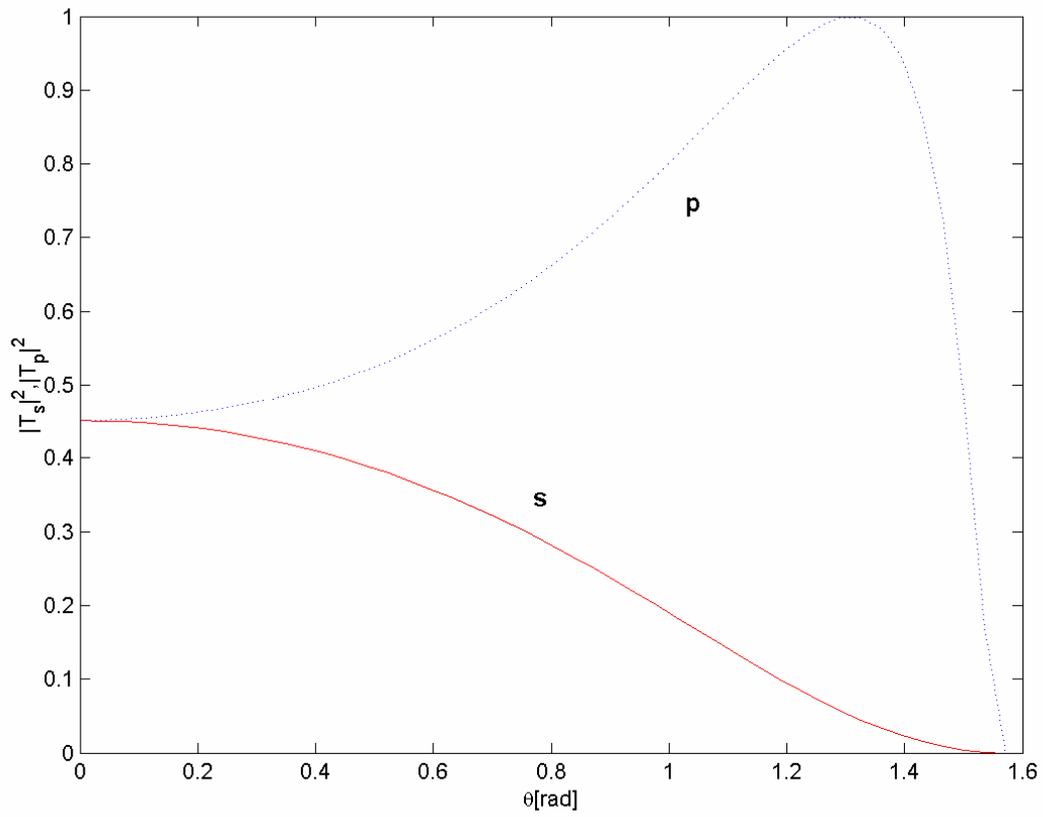

**FIGURE 3a :**

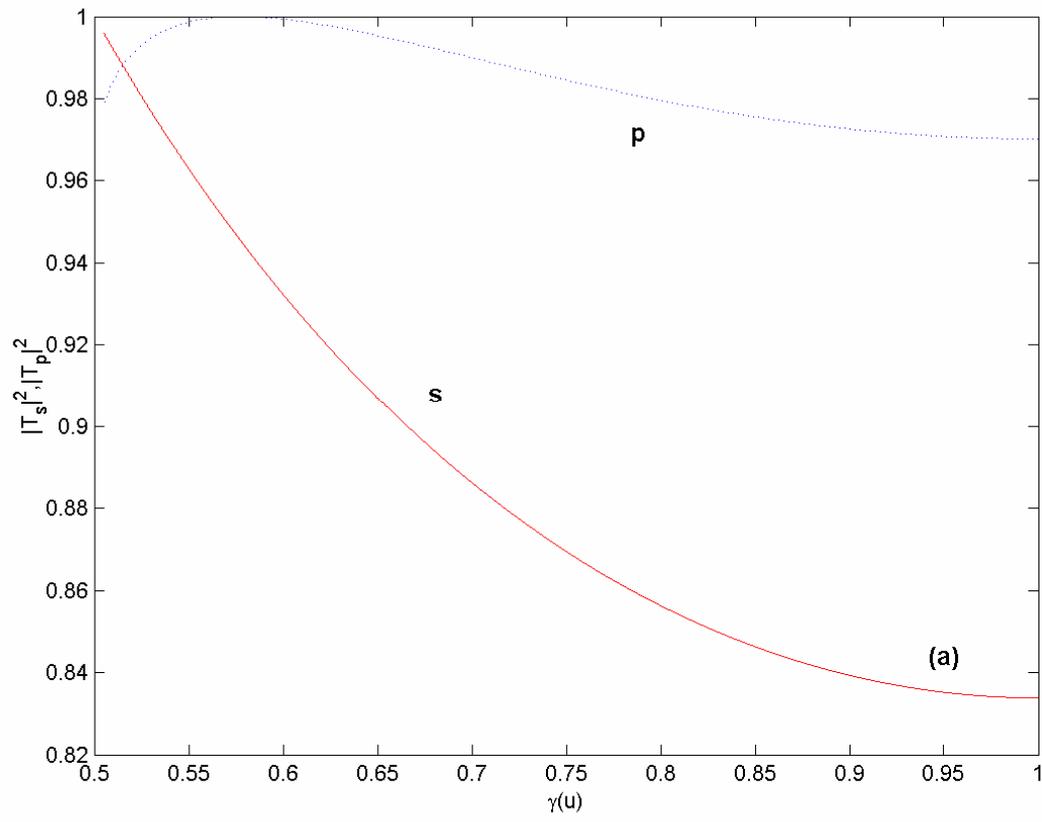

**FIGURE 3b:**

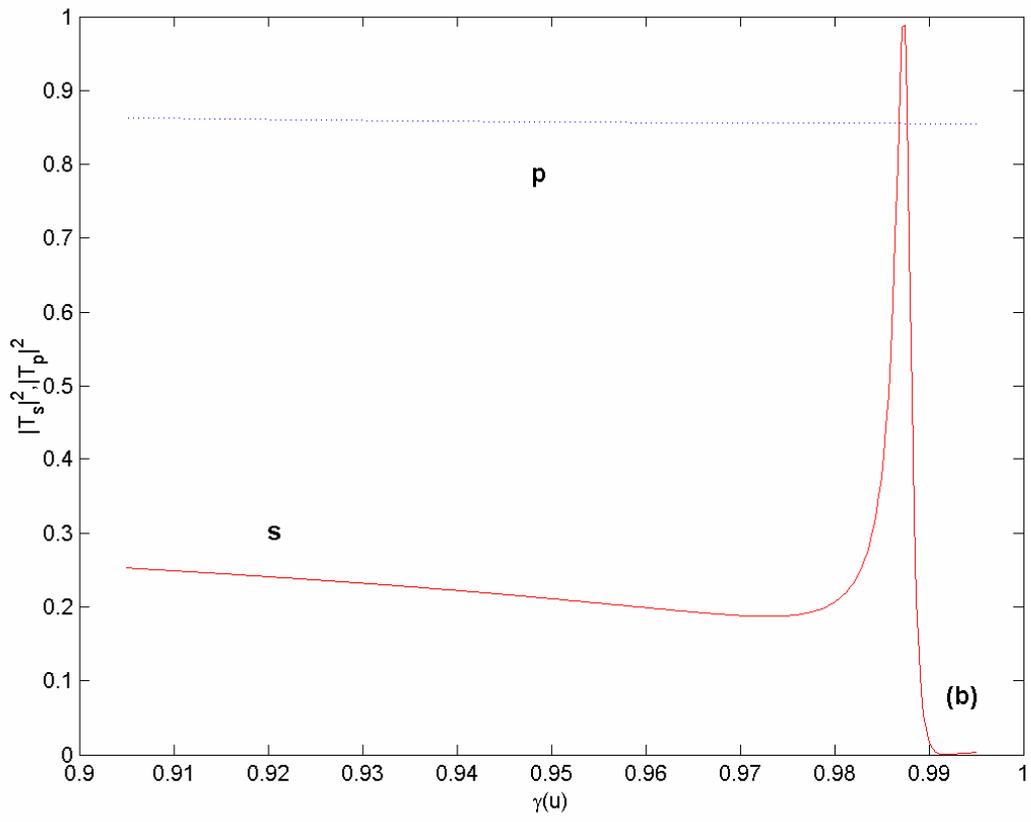

**FIGURE 3c :**

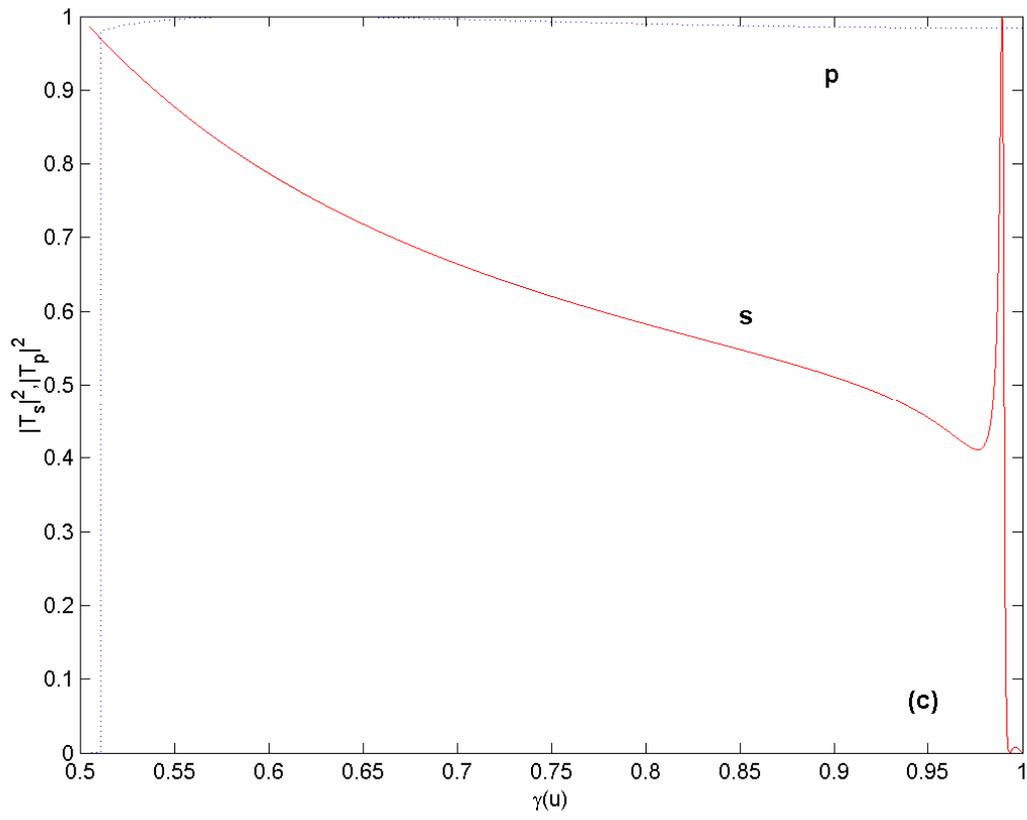

**FIGURE 4a :**

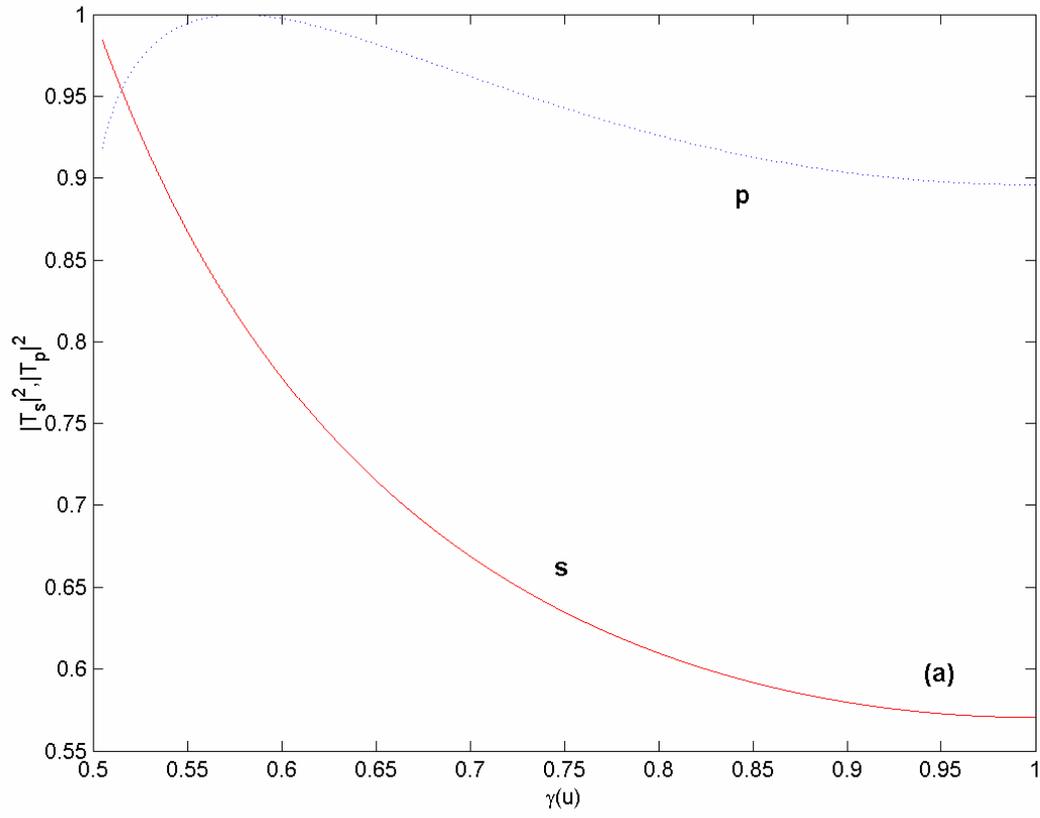

**FIGURE 4b :**

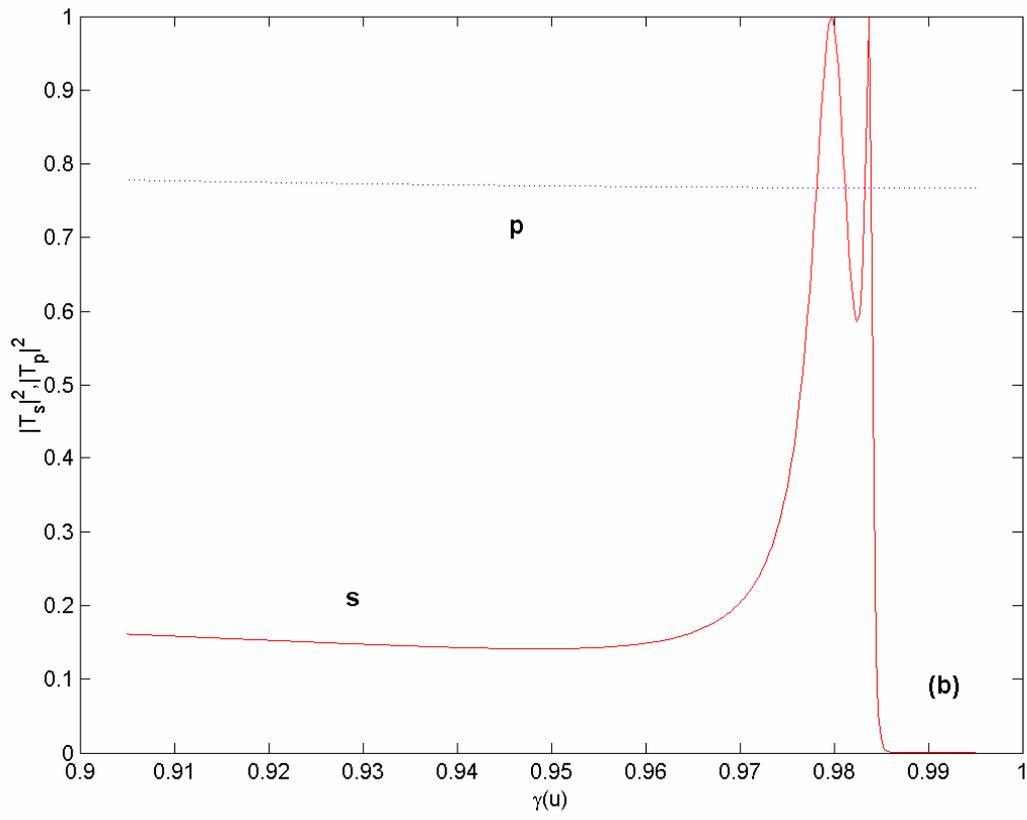

**FIGURE 5 :**

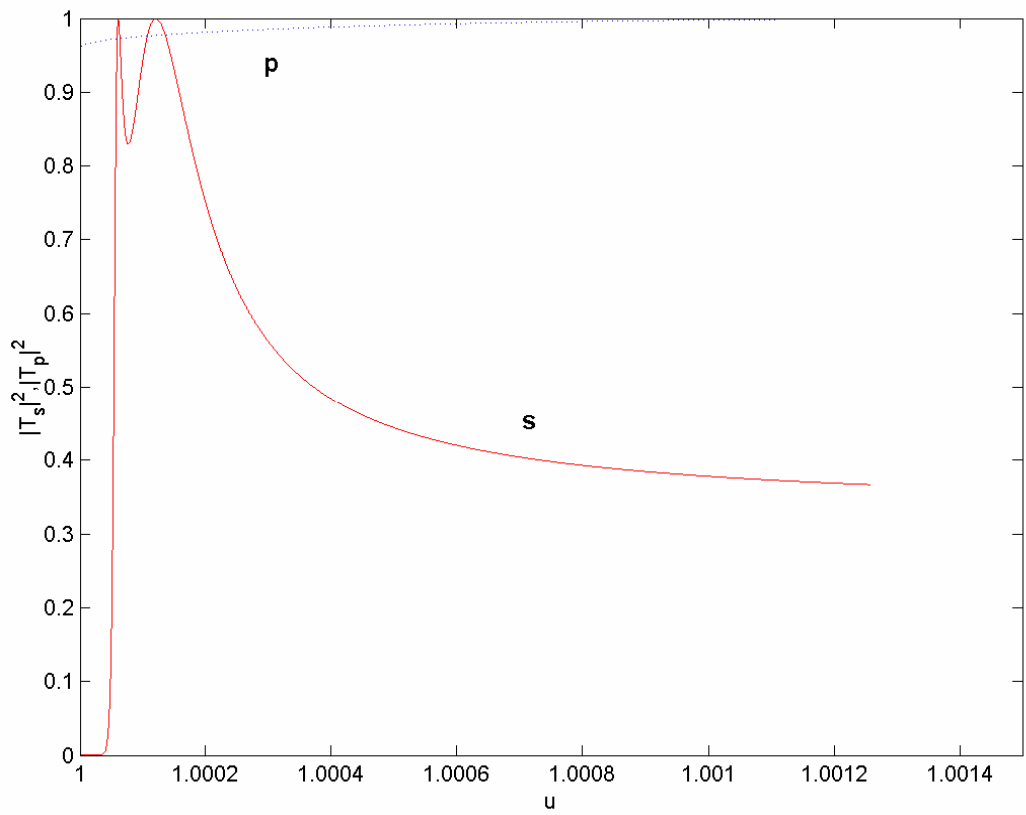

**FIGURE 6 :**

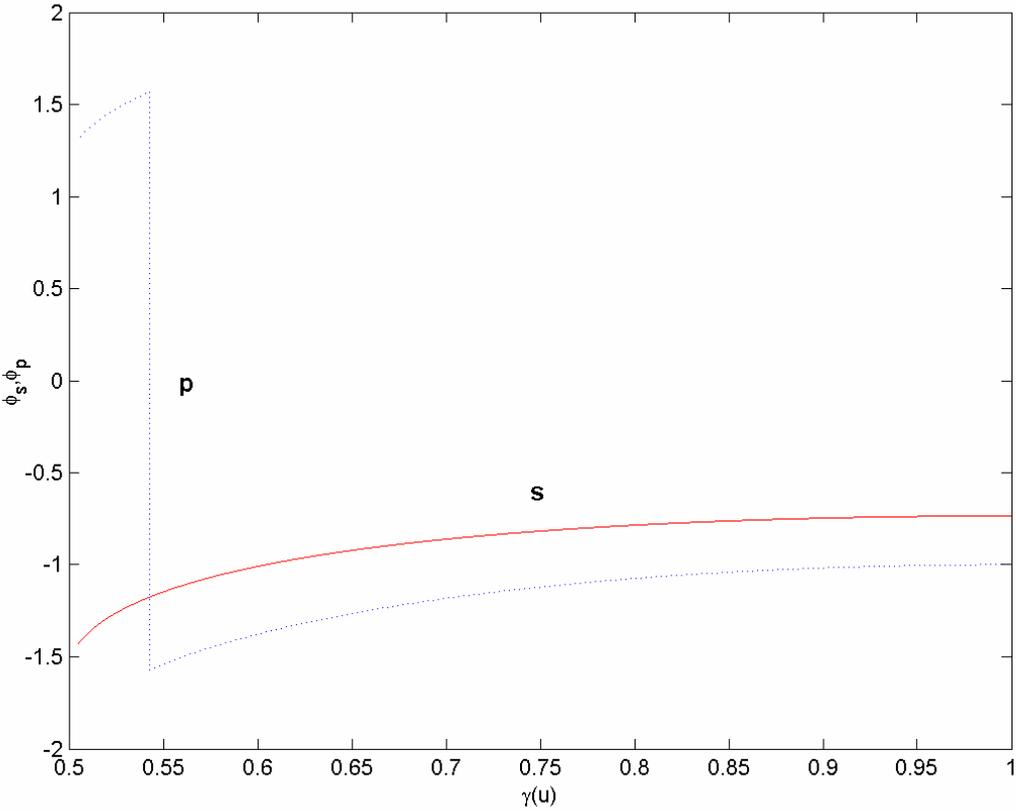

**FIGURE 7 :**

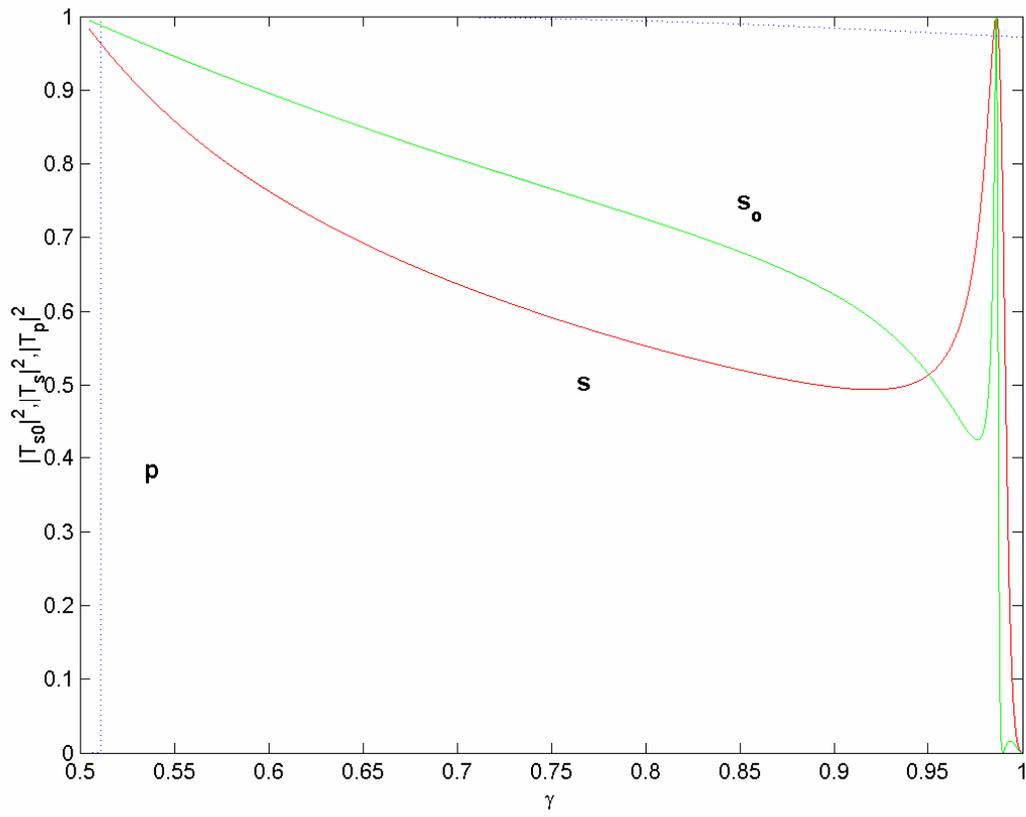